\documentclass[prd,aps,showpacs,nofootinbib,preprint]{revtex4-1}
\usepackage{graphicx,color,amsmath,amsxtra}
\usepackage{epsf}
\usepackage{amssymb}
\usepackage{enumerate}
\usepackage{hhline}
\usepackage{array}
\usepackage{tabularx}
\usepackage{hangcaption}

\newcommand{\newc}{\newcommand}
\newc{\be}{\begin{equation}}
\newc{\ee}{\end{equation}}
\newc{\beq}{\begin{eqnarray}}
\newc{\eeq}{\end{eqnarray}}

\begin{document}
\rightline{\tt{MIT-CTP-4272}}

\title{Constrained Jackiw-Rebbi model gives McGreevy-Swingle model
}
\author{
S.-H. Ho}
\affiliation{Center for Theoretical Physics \\
Massachusetts Institute of Technology \\
 Cambridge, MA 02139}


\begin{abstract}
We show that the recently considered McGreevy-Swingle model for Majorana fermions in the presence of a 't Hooft-Polyakov magnetic monopole arises when the Jackiw-Rebbi model is constrained to be conjugation self dual.
\end{abstract}

\maketitle

The Dirac equation in a topological background has been studied in various dimensions, such as the background of a kink in one spatial dimension, a vortex in two spatial dimensions, a 't Hooft-Polyakov magnetic monopole and a dyon in three dimensions \cite{Jackiw:1975fn}\cite{Jackiw:1981ee}, and there exist normalizable Dirac zero modes in all cases. The zero modes of Majorana fermions, however, are only found in the cases of a kink in one dimension \cite{Jackiw:1975fn} and a vortex background in two dimensions \cite{Jackiw:1981ee}. 
A well separated pair of Majorana zero modes can define a quibit since it is a degenerate two-state system, whose state is stored nonlocally \cite{Hasan:2010xy}. Due to this feature and obeying non-Abelian statistics, Majorana zero modes caught a lot of attention in physics because of its potential application on quantum computing \cite{Nayak:2008zza}.


Recently, McGreevy and Swingle considered a three spatial dimension model for Weyl fermions coupled to a 't Hooft-Polyakov monopole and a scalar field in the $SU(2)$ adjoint representation \cite{McGreevy:2011if} \cite{Teo:2009qv}. In \cite{McGreevy:2011if}, they solved the zero mode Dirac equation explicitly and found the exact solutions for the Majorana zero modes. In this brief report, we show that the Jackiw-Rebbi model with a Dirac fermion in the fundamental representation of $SU(2)$ gauge group, once the conjugation condition is imposed on the Dirac field, reproducs the single Wyel fermion case of McGreevy-Swingle model. This indicates that the Majorana feature of the model is not only shown in the zero mode, but in the whole field. However, 
the quantum version of the theory is problematic because of the Witten anomaly \cite{McGreevy:2011if}\cite{Witten:1982fp}.

Let us start from the Lagrangian density (3.1) in \cite{Jackiw:1975fn}:

\begin{subequations}
\beq \label{eq1}
\cal{L} && = \bar{\psi}_a i \gamma^{\mu} (D_{\mu})_{ab} \psi_b -gG \bar{\psi}_a T^A_{ab} \psi_b \Phi^A , \\
(D_{\mu})_{ab} && =\partial_{\mu} \delta_{ab}-i g A_{\mu}^A T^A_{ab}  
\eeq
where $\psi_a$ is a four-component Dirac spinor and a two-component $SU(2)$ isospinor, $A_{\mu}^A$ is the vector potential, $T^A$ is the $SU(2)$ generator, $g_{\mu\nu}=diag (+1, -1, -1, -1) $ and $G$ is a positive dimensionless coupling constant. Here $(a,b)$ are isospin indices while spin indices are suppressed. (We work in the chiral representation for the gamma matrices and the fundamental representation for the $SU(2)$ matrices. Thus we use gamma matrix conventions of \cite{McGreevy:2011if} rather than \cite{Jackiw:1975fn}.)

\beq \label{eq2}
\gamma^{\mu} && = \left(\begin{array}{cc}0 & \sigma^{\mu} \\\bar{\sigma}^{\mu} & 0\end{array}\right), \\
T^{A} && =\frac{\tau^A}{2}, \ A=1,2,3
\eeq
Here $\sigma^{\mu}=(1, \vec{\sigma})$ and $\bar{\sigma}^{\mu}=(1, -\vec{\sigma})$. 
\end{subequations}

From (\ref{eq1}) we can derive the Dirac equation
\beq \label{eq3}
\left(i\gamma^{\mu} (\partial_{\mu} \delta_{ab}-i  \frac{g}{2} A_{\mu}^A \tau^A_{ab}) - \frac{gG}{2}  \tau^A_{ab} \Phi^A \right) \psi_b =0
\eeq
or equivalently
\begin{subequations}
\beq \label{eq4}
&&H_{ab} \psi_b \equiv \left[ \vec{\alpha} \cdot \vec{p} \delta_{ab} + \frac{g}{2} \vec{\alpha} \cdot \vec{A^A} \tau^A_{ab} +\frac{g\ G}{2} \beta \tau^A_{ab} \Phi^A \right] \psi_b = i \partial_t \psi_a= E \psi_a, \\
&& \vec{p}=-i \vec{\nabla}, \\
&&\vec{\alpha}=\gamma^0 \vec{\gamma}=\left(\begin{array}{cc}-\vec{\sigma} & 0 \\0 & \vec{\sigma}\end{array}\right), \ \ 
\beta =\gamma^0 =\left(\begin{array}{cc}0 & 1 \\1 & 0\end{array}\right).
\eeq

\end{subequations}

The conjugated field 
\beq  \label{eq5}
\psi_a^c \equiv \left(\begin{array}{cc}0 & i \sigma^2 \\i \sigma^2 & 0\end{array}\right)  i \tau^2_{ab} \psi^*_b \equiv C_{ab} \psi^*_b
\eeq
satisfies the equation 
\beq  \label{eq6}
H_{ab}\psi_b^c = -E \psi_a^c 
\eeq
owing to  
\beq  \label{eq7}
(C H C^{-1})_{ab} = - (H^*)_{ab} .
\eeq

Now we impose the conjugation constraint on the Dirac spinor $\Psi_a = \left(\begin{array}{c}\xi_a \\\eta_a\end{array}\right) $,
\begin{subequations}
\beq \label{eq8}
\Psi_a^c && = \Psi_a, \\
\xi_a = i \sigma^2 i \tau^2_{ab} \eta_b^*  &&, \ \ \eta_a =i \sigma^2 i \tau^2_{ab} \xi_b^* .
\eeq
\end{subequations}

Replacing the unconstrained Dirac spinor $\psi_a$ by the constrained $\Psi_a$ we can rewrite (\ref{eq1}) in terms of the two component field $\xi_a$:
\beq \label{eq9}
\cal{L} && = \Psi^{\dag}_a \left(\begin{array}{cc}i \bar{\sigma}^{\mu} \left( \partial_{\mu} \delta_{ab} - \frac{ig}{2} A_{\mu}^A \tau^A_{ab} \right) & -\frac{gG}{2} \Phi^A \tau^A_{ab} \\-\frac{gG}{2} \Phi^A \tau^A_{ab} & i \sigma^{\mu} \left( \partial_{\mu} \delta_{ab} - \frac{ig}{2} A_{\mu}^A \tau^A_{ab} \right)\end{array}\right) \Psi_b \\
&&= 2 \xi^{\dag}_a i \bar{\sigma}^{\mu} \left( \partial_{\mu} \delta_{ab} - \frac{ig}{2} A_{\mu}^A \tau^A_{ab} \right) \xi_b -\frac{gG}{2} \xi_a^{T} ( i\tau^2  \tau^A)_{ab} \Phi^A i \sigma^2 \xi_b - \frac{gG}{2} \xi^{\dag}_a (i \tau^A \tau^2)_{ab} \Phi^A i \sigma^2 \xi^*_b \nonumber \\
\eeq
This is Equation (2.1) in \cite{McGreevy:2011if}. The single zero mode $\psi_a^0$ is present both in the unconstrained JR model and the constrained McGS model since its mode function satisfies $\psi_a^0 = C_{ab} \psi_b^{0*}$. 

A similar story has been told in two spatial dimensions: an unconstrained Dirac equation with conjugation properties like (\ref{eq5}) and (\ref{eq6}) describes graphene; when a conjugation constraint is imposed, the equation reduces to two components and describes Majorana fermions \cite{Chamon:2010ks}.

We have shown that the McGS model emerges when the energy reflection conjugation 
is imposed on JR model with a Dirac fermion in $SU(2)$ fundamental representation. Here we also note that the Majorana fermion in $SU(2)$ adjoint representation in JR model cannot be achieved since it seems to be no way to impose the 
conjugation constraint in isovector fermion case \footnote{In the case of isovector fermion, the corresponding conjugated field is defined by
$\Psi^c_a \equiv \left(\begin{array}{cc}0 & i \sigma^2 \\i \sigma^2 & 0\end{array}\right) \Psi^*_a$. Once we impose the conjugation constraint $\Psi^c=\Psi$ we only have the trivial solution $\Psi=0$.}.

\begin{acknowledgements}
We thank R. Jackiw for suggesting this calculation and for discussion with J. McGreevy and B. Swingle. This work is supported by the National Science Council of R.O.C. under Grant number:
NSC98-2917-I-564-122.

\end{acknowledgements}

\end{document}